\documentclass[]{aa}
\usepackage{graphicx}
\usepackage{natbib}
\usepackage{times}
\bibpunct{(}{)}{;}{a}{}{,} % to follow the A&A style
\let\bibfont=\small%tiny

\begin{document}
  \title{Evolution of clouds in radio galaxy cocoons}

  \author{G. Mellema \and J. D. Kurk \and H. J. A. R\"ottgering}

  \institute{Sterrewacht Leiden, P.O. Box 9513, 2300 RA, Leiden, 
             The Netherlands}

  \offprints{G.~Mellema\hfill\\ \email{mellema@strw.leidenuniv.nl}}

  \date{Received 31 July 2002 / Accepted 6 September 2002}

  \abstract{This letter presents a numerical study of the evolution of
  an emission line cloud of initial density 10~cm$^{-3}$,
  temperature $10^4$~K, and size 200~pc, being overtaken by a strong
  shock wave. Whereas previous simple models proposed that such a
  cloud would either be completely destroyed, or simply shrink in
  size, our results show a different and more complex behaviour: due
  to rapid cooling, the cloud breaks up into many small and dense
  fragments, which can survive for a long time.  We show that
  such rapid cooling behaviour is in fact expected for a wide range of
  cloud and shock properties.

  This process applies to the evolution of emission line clouds being
  overtaken by the cocoon of a radio jet. The resulting small clouds would be
  Jeans unstable, and form stars. Our results thus give theoretical
  credibility to the process of jet induced star formation, one of the
  explanations for the alignment of the optical/UV and radio axis observed in
  high redshift radio galaxies.

   }

  \maketitle
%
%________________________________________________________________

\section{Introduction}

Regions of luminous optical line and continuum emission near high redshift
radio galaxies ($z > 0.6$) are often found to be extended along the direction
of the radio axis \citep{Chambers87,McCarthy87}. One obvious explanation for
these alignments, is that star formation takes place in regions where the
shock bounding the radio jet, has passed.

Recent observations seem to support this idea. Deep spectra of the radio
galaxy 4C41.17 at $z = 3.8$, show that the bright, spatially extended
rest-frame UV continuum emission is unpolarized and contains P Cygni-like
absorption features, indicating the presence of a large population of young,
hot stars \citep{Deyetal97}. \citet{bicknelletal00} argue that this can best
be understood if the shock associated with the radio jet has triggered star
formation within the emission line clouds.

A nearby example where stars might be formed under the influence of a
radio source is the case of Cen A. Here, young stars are found near
filaments of ionized gas in a radio lobe \citep{mouldetal2000}.

\citet{Rees89} and \citet{Begelmanetal89} analytically explored the evolution
of intergalactic medium (IGM) clouds, overtaken by shocks from the cocoon of a
radio jet. They argue that these clouds would be compressed and then
gravitationally contract to form stars. However, \citet{Icke99} claimed that
the destructive aspects of the interaction between the expanding cocoon and
the clouds would dominate the evolution of the clouds. In his scenario the
clouds evaporate and their material mixes into the jet cocoon.

Given the complexity of the interaction between the clouds and the jet cocoon,
numerical studies are a good tool to investigate this problem. Although the
`shock-cloud interaction' problem was studied numerically before, none of
these studies addresses the effects of radiative cooling, important for
intergalactic clouds. Here we present new results of a numerical hydrodynamic
study of the shock-cloud interaction problem, including the effects of
radiative cooling.

In Sect.~2 we describe the general problem of shock-cloud interaction and the
application to IGM clouds. Section~3 deals with the numerical method, and
Sect.~4 contains the results, which we further discuss in the fifth
section. We sum up the conclusions in Sect.~6.
%__________________________________________________________________

\section{Shock-Cloud Interactions}
Many numerical studies of single shock-cloud interactions have been carried
out, \cite{Woodward76} being one of the first. Various others followed, of
which we will only mention two more recent studies: \cite{Kleinetal94}, who
provided a thorough analysis of the problem, and \cite{Poludnenkoetal01}, who
studied the case of a shock running over a system of clouds; see these two
papers for an overview of the literature. It is notable that in nearly all
numerical studies to date, radiative cooling was either neglected or had
little effect. For work considering the large scale effects of the
passage of radio jets, see \citet{Steffenetal97} and \citet{Reynoldsetal01}.

The evolution of a single, non-cooling cloud, which is run over
by a strong shock wave, consists of three phases. Initially, the shock runs
over the cloud. The time scale for this is the shock passing time, $t_{\rm sp}
= 2R_{\rm cl}/v_{\rm shock}$, where $R_{\rm cl}$ is the cloud radius, and
$v_{\rm shock}$ the velocity of the passing shock.

The second phase is the compression phase, in which the cloud finds itself
inside the high pressure cocoon.  It is now underpressured compared to its
environment, and shock waves start to travel into the cloud from all
sides. This phase lasts for a time $t_{\rm cc}=R_{\rm cl}/v_{\rm s,cl}$, the
cloud crushing time, where $v_{\rm s,cl}$ is the velocity of the shock
travelling into the cloud. For a strong shock this velocity is of order
$v_{\rm s,cl}=v_{\rm shock}/\sqrt{\chi}$, in which $\chi$ is the ratio of the
$n_{\rm cl}$ to $n_{\rm env}$, the densities of the cloud and the environment,
respectively; see \citet{Kleinetal94} for a better estimate.

The third phase starts when the shocks travelling into the cloud, meet and
interact. This produces a rarefaction wave travelling through the shocked
cloud material. The cloud, which was compressed by the shock waves, now starts
expanding again, and soon afterwards is destroyed and mixes in with the
surrounding flow. This typically happens in a few cloud crushing times.

\subsection{Cloud properties}

Following \citet{Rees89}, \citet{Begelmanetal89}, and \citet{McCarthy93},
we assume the undisturbed clouds to be the cooler and denser phase of an
ionized two-phase IGM, of which the low density phase has a temperature of
$T_{\rm ig}=10^7$~K and a density of $n_{\rm ig}=10^{-2}$~cm$^{-3}$.  Assuming
pressure equilibrium between the two phases, a cloud temperature of $10^4$~K
gives a density of $n_{\rm cl}=10$~cm$^{-3}$.  We choose an initial radius of
$3\times 10^{20}$~cm ($\sim 100$~pc) and hence the cloud mass is $9.5\times
10^5$~$M_\odot$.  Following the analysis of Cygnus~A by
\citet{Begelmanetal89}, we take the Mach number of the shock bounding the jet
cocoon to be 10, yielding $v_{\rm shock}$= 3500~km~s$^{-1}$ (0.01$c$). With
these parameters we obtain $t_{\rm sp}=5\times 10^4$~years, $t_{\rm
cc}=8\times 10^5$~years, and $v_{\rm cl,s}=120$~km~s$^{-1}$.

The cooling time can be estimated from $t_{\rm cool}=Cv_{\rm s,cl}^3/\rho_{\rm
cl}$ \citep[see e.g.][]{Kahn76}, where $C$ is a constant depending on the
cooling processes, with a value of $6.0\times 10^{-35}$~g~cm$^{-6}$~s$^4$ for a
gas in collisional ionization equilibrium at solar abundances.  With the
numbers above one finds $t_{\rm cool}=2\times 10^2$~years. This is the shortest
time scale thus far, showing that cooling will dominate the evolution of the
shocked cloud.

It is instructive to derive a condition for which cooling will dominate.
Using the expressions for $t_{\rm cc}$ and $t_{\rm cool}$, we find that 
the condition $t_{\rm cc}>10t_{\rm cool}$ can be rewritten as 
\begin{eqnarray}
 M_{\rm cl} &>& 10^{-9} M_\odot \times\nonumber\\
&&
\left({v_{\rm shock}\over 10^3 {\rm km~s}^{-1}}\right)^{12}
\left({\chi \over 10^3}\right)^{-8}
\left({n_{\rm e}\over 10^{-2} {\rm cm}^{-3}}\right)^{-2}\,.
\end{eqnarray}
This shows that cooling dominates for a large range of values for
$M_{\rm cl}$, $v_{\rm shock}$, $\chi$, and $n_{\rm e}$. For our
values of $M_{\rm cl}$ and $n_{\rm e}$, the shock velocity needs to be
above 17\,000~km~s$^{-1}$, or the density ratio $\chi$ below 15, for
cooling {\it not}\/ to dominate the evolution.

\begin{figure*}
\centerline{\includegraphics[height=79mm,clip=]{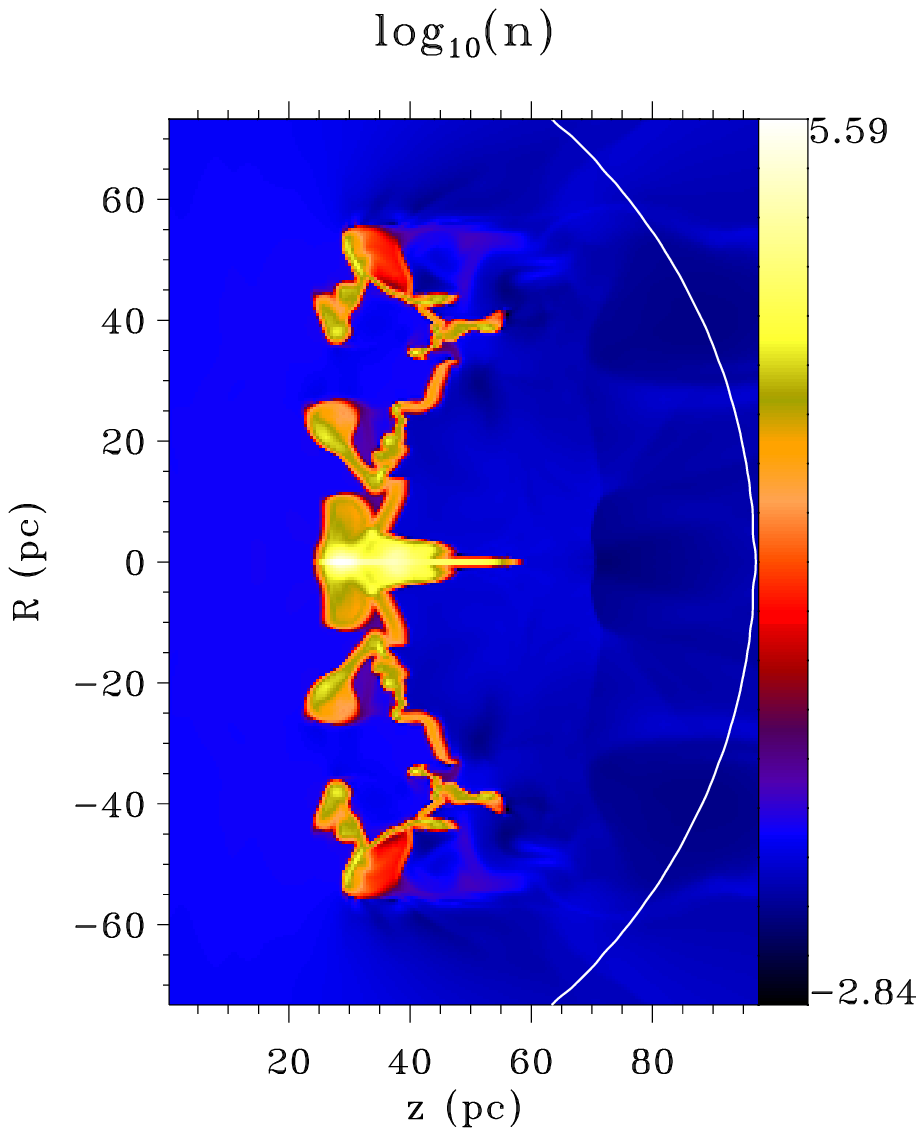}
\includegraphics[height=79mm,clip=]{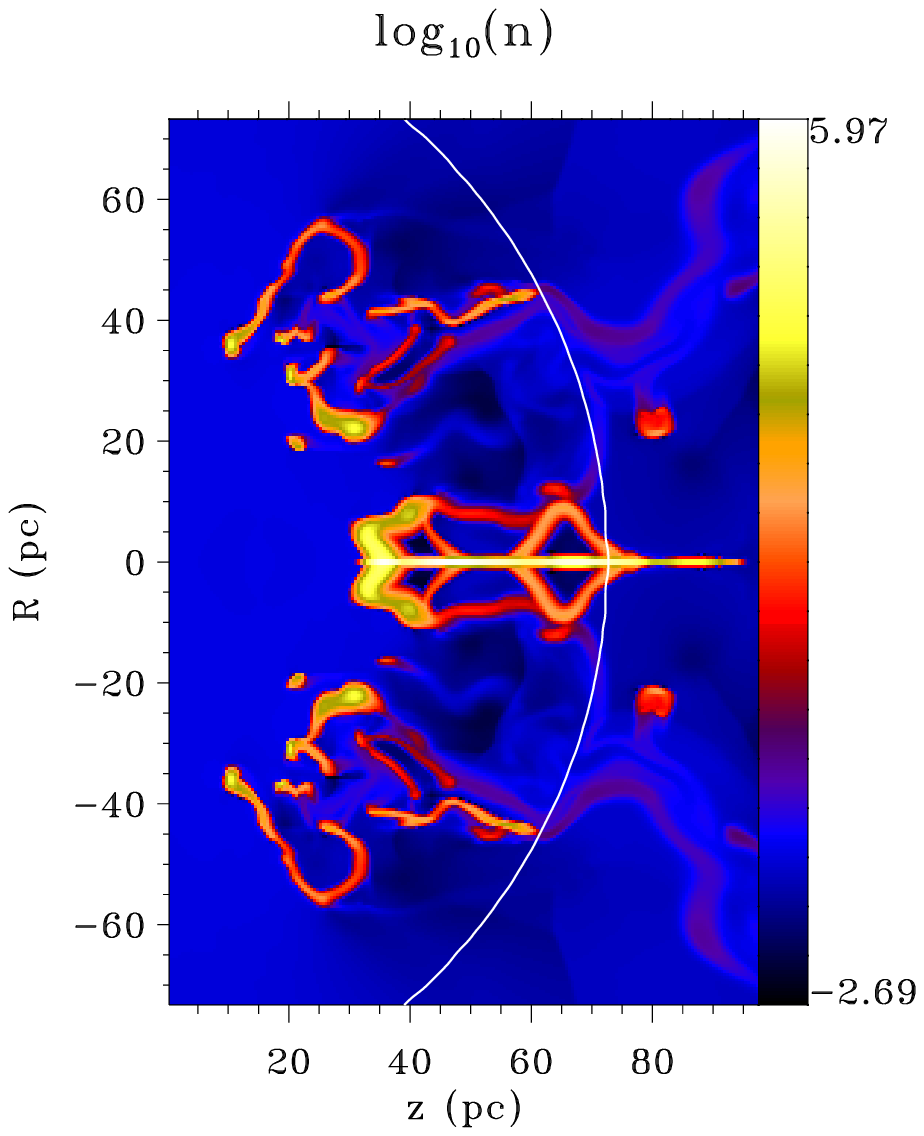}}
\caption{Colour plots of ${\log}_{10}$ of the number density (cm$^{-3}$) at
 $t=0.79\times 10^6$~years (left), and $t=1.1\times 10^6$~years (right) for
 run A (cylindrical coordinates, spherical cloud). The white contour indicates
 the original cloud position. Only a fraction of the computational
 domain is shown.  The coordinates are relative to the lower left corner of
 each frame. The shock wave came from the left.}
\label{logdensA}
\end{figure*}

\begin{figure*}
\centerline{\includegraphics[height=79mm,clip=]{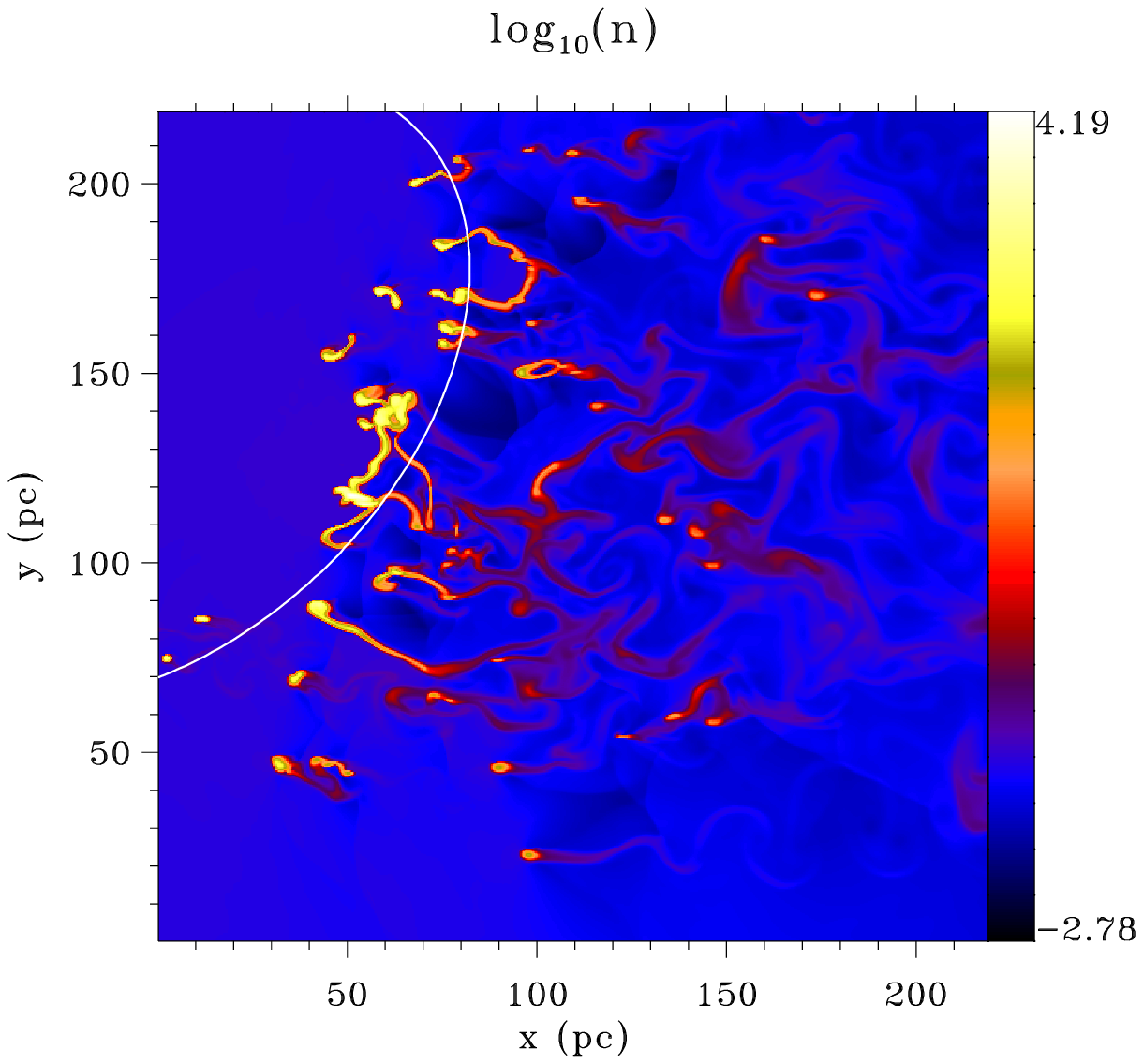}
\includegraphics[height=79mm,clip=]{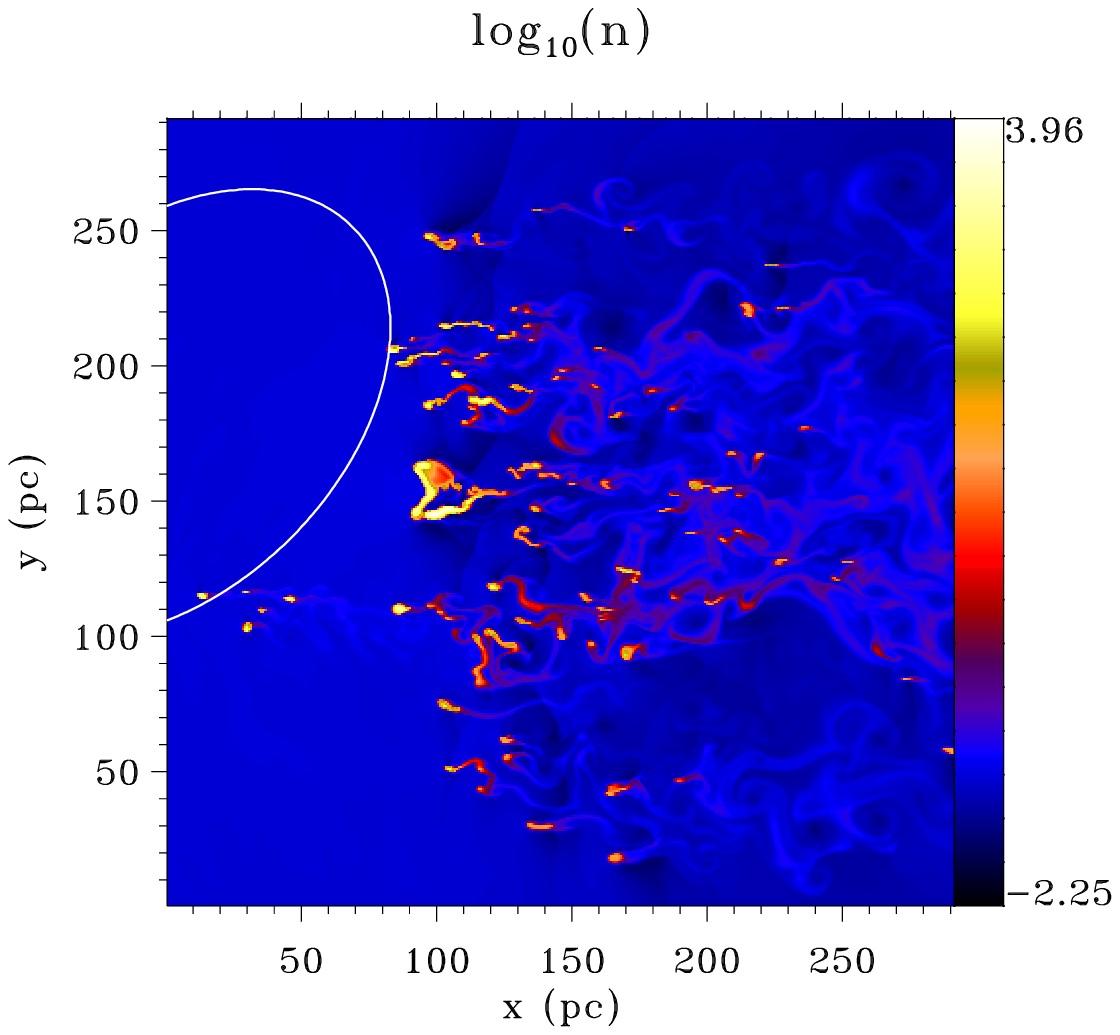}}
\caption{As Fig.~1 for run B (cartesian coordinates, elliptical cloud).}
\label{logdensB}
\end{figure*}

\section{Numerical method}

The calculations were performed with a two-dimensional hydrodynamics code based
on the Roe solver method, an approximate Riemann solver \citep{Roe81,EulMel}.
Second order accuracy was achieved with the {\it superbee}\/ flux limiter,
which was made less steep by lowering the coefficients from 2.0 to 1.2; taking
1.0 would correspond to using the {\it minmod}\/ flux limiter, see Sect.~20.2
in \citet{laney}. Better two-dimensional behaviour was implemented by using
the transverse waves method as described by \citet{LeVeque}.

In order to include the effects of cooling, we used a cooling curve
\citep{DalgarnoMcCray}, which gives the cooling as function of
temperature, for a low density plasma in collisional ionization
equilibrium. This is a reasonable approximation of the real cooling
processes of astrophysical gases.  The radiative terms were
implemented using operator splitting, where the appropriate radiative
losses and gains were added as a separate source term every time
step. The heating rate is proportional to the density, and was set so
that for the initial conditions, heating and cooling in the cloud are
balanced.

In order to deal with short cooling times, we subdivided the time
steps into smaller fractions of the order of the cooling time when
applying the cooling. We imposed a minimum temperature of 10~K. This
approximately corresponds to the cosmic microwave background
temperature at the redshifts we are considering.  We did not follow
the ionization state of the gas, but assumed the material to always be
in collisional ionization equilibrium.

The geometry of the grid was either cylindrical $(R,z)$, assuming cylindrical
symmetry, or cartesian $(x,y)$, assuming slab symmetry. The use of two
different coordinate systems helps in understanding the true
three-dimensional nature of the flow. Cylindrical coordinates are the proper
choice as long as the flow pattern retains its large scale character, {i.e.\
}during the initial phase of the interaction. However, when the cloud starts to
fragment, off-axis pieces are represented by ring-shaped structures.
Furthermore, there is a strictly imposed symmetry axis at the centre of the
cloud. In cartesian coordinates the initial conditions do not describe a
spheroid, but rather a cylinder. On the other hand, the fragmentation is more
properly followed, and no symmetry axis is imposed.

We ran two simulations: in run A the shock wave interacted with a spherical
cloud (with the parameters from Sect.~2.1) on cylindrical coordinates, and in
run~B with an elliptical cloud (with a semi-major axis of 100~pc, axis ratio
1.5, the major axis at an angle of $45^\circ$ with respect to the incoming
shock, and all other properties the same as in run A) on cartesian
coordinates. Using an elliptical cloud, rather than a spherical cloud, further
reduces the symmetry. For both runs the cell sizes were $0.486\times
0.486$~pc, using $800\times 1600$ (A) and $1600\times 1600$ (B) computational
cells.

\section{Results of the simulations}

Figure~\ref{logdensA} shows the logarithm of the density for run A at times
$0.79\times 10^6$ and $1.1\times 10^6$~years. Figure~\ref{logdensB} shows the
same for run B\footnote{Movies of the entire density evolution of the two runs
are available with the electronic version of this letter.}. The cocoon shock
wave came from the left, and passed the entire cloud at $t=5\times
10^4$~years. At $t=0.79\times 10^6$~years the shock waves travelling into the
cloud have just merged (compare with the estimate for $t_{\rm cc}$ in
Sect.~2.1). In the non-cooling case this is followed by a re-expansion of the
shocked cloud (due to the extra heating generated in the merging of the
shocks), but here the excess energy is radiated away, and the merging of the
front- and back-side shocks leads to the formation of a dense, cool,
elongated, but fragmented structure (`sheet') perpendicular to the flow
direction in run~A, and more parallel to the major axis orientation in
run~B. In both cases there is a concentration near the centre of the former
cloud.

The two righthand boxes of Figs.~\ref{logdensA} and \ref{logdensB} show how
this sheet fragments further. In run A, the imposed symmetries lead to an
elongated concentration of material on the axis, which we measured to contain
approximately 10\% of the original cloud mass. The rest of the cloud material
is compressed into dense structures, spread out over a volume which is 30\%
of the original cloud size (part of the outer contour of the original cloud 
boundary is indicated in Fig.~\ref{logdensA}).

In run B the cloud develops into an ensemble of dense small fragments,
filling an area of approximately the same diameter as the
original cloud. Without imposing symmetry, the largest and densest
fragment is found near the centre of the ensemble. The integrated
density of this largest fragment was measured to be some 30\% of the
integrated density of the original cloud.

In both runs, at the end of the simulation, less than a percent of the
original cloud material has been mixed into the cocoon, showing that the
evaporation process is slow, as is expected for high density contrasts. The
ensemble does spatially disperse since the velocities of the fragments
range from 90 to 500~km~s$^{-1}$, the leftmost fragments having the lowest
velocities.

\section{Discussion}
The simulations presented here show a completely new behaviour
compared to the scenarios presented in \citet{Rees89} (compression),
or \citet{Icke99} (disruption). In our simulations, instead of being
simply compressed or disrupted, the cloud breaks up into many small
dense fragments, spread out over a certain volume, and which evaporate
only slowly. This has not been seen in numerical simulations
before. It is completely due to the introduction of cooling, which
prevents the `third phase' or re-expansion (see Sect.~2).

We expect that full three-dimensional simulations will show a result lying
somewhat in between what we found in runs A and B. It would definitely enhance
rather than suppress the fragmentation, since there is one extra degree of
freedom available for instabilities \citep[see e.g.][]{xu&stone}.

There are a number of processes which could work against the cooling, and
hence slow down the compression. These are for example heating by the UV and
X-ray photons from the AGN and the presence of a magnetic field in the clouds.
Simulations of magnetized flows in three dimensions, as reported by
\citet{gregorietal}, show that if the magnetic field is strong enough, it will
actually enhance the fragmentation of the cloud, and presumably aid
evaporation rather than compression. However, these simulations did not
include the effects of cooling, so it is difficult to compare their results to
ours.

Note that whenever the cooling time is substantially shorter than the
cloud crushing time, we expect an evolution similar to the one above.
Equation 1 shows that this holds for a wide range of cloud parameters. For
example, the {\it interstellar} clouds from \cite{Poludnenkoetal01} should
strongly cool, be compressed, and develop into a long-lived mass loading flow,
something which these authors failed to achieve in their non-cooling
simulations, where the clouds are destroyed within a few $t_{\rm cc}$.

The further evolution of our fragments will be dominated by two processes:
gravitational collapse, and further acceleration and erosion by the passing
flow. All fragments found in our simulations will collapse under their own
gravity, which makes them smaller, and even harder to disrupt and/or
accelerate. As pointed out in Sect.~4, nearly all of the original cloud
material ends up in these dense fragments, and would be available for star
formation. This implies that the estimate for the induced star formation rate
from \citet{Begelmanetal89}, is still valid. For a cloud filling factor (by
volume) of $10^{-3}$, and a relativistic jet, they find an induced star
formation rate of $\sim 100$~M$_\odot$~yr$^{-1}$, in rough agreement with the
observations.

\section{Conclusions}
We have for the first time simulated the cooling dominated evolution of an
intergalactic cloud which is overrun by the cocoon of a passing radio
jet. Previous analytical studies conjectured that the cloud would either be
compressed, or be completely destroyed and evaporate into the cocoon. We
instead find a new picture. Radiative cooling is so rapid, that nearly all of
the cloud mass is compressed into many small and dense fragments with a long
hydrodynamical survival time. These fragments are likely to collapse and form
stars, in line with the scenario of jet induced star formation.

This type of fragmentation is expected whenever the cooling time is much
shorter than the cloud crushing time. Evaluating this condition, shows this to
be case for a wide range of parameters, stretching from intergalactic to
interstellar conditions, see Eq.~1. The collapse-and-fragment sequence we
find, may well be the way to create long lived mass loading flows inside
post-shock regions \citep{HartquistDyson}. 

These simulations are only a first step, and definitely more work is needed.
In future papers we plan to explore the effects three-dimensionality and
self-gravity have on the fragmentation process.

\begin{acknowledgements}
This work was sponsored by the National Computing Foundation (NCF) for the 
use of supercomputer facilities, with financial support from the Netherlands 
Organization for Scientific Research (NWO).

The research of GM has been made possible by a fellowship of the Royal
Netherlands Academy of Arts and Sciences.  
\end{acknowledgements}

\bibliographystyle{aa}
\bibliography{special.bib}

\end{document}